\documentclass{article}
\usepackage{amsfonts}
\usepackage{amsmath}
\usepackage{hyperref}
\usepackage{epsfig}

\usepackage[OT2,OT1]{fontenc}
\newcommand\cyr{%
  \renewcommand\rmdefault{wncyr}%
  \renewcommand\sfdefault{wncyss}%
  \renewcommand\encodingdefault{OT2}%
  \normalfont
  \selectfont}
\DeclareTextFontCommand{\textcyr}{\cyr}

\def\Dir{\,\,{\raise.15ex\hbox{/}\mkern-12mu D}}

\newcommand{\bb}{\begin{equation}}
\newcommand{\bqn}{\begin{eqnarray}}
\newcommand{\eqn}{\end{eqnarray}}

\def\({\left(}
\def\){\right)}
\def\[{\left[}
\def\]{\right]}

\begin{document}

\vspace*{1cm}

\begin{center}
{\bf{\Large A (non)static 0-order statistical model\\
and its implementation for compressing virtually uncompressible data }}

\vspace*{1cm}

Evgeniy Vitchev\footnote{e-mail: vitchev@physics.rutgers.edu}
\ \\

\ \\
Department of Physics and Astronomy, Rutgers University \\
Piscataway, NJ 08855-0849, USA
\end{center}

\vspace*{.8cm}

\begin{abstract}
We give an implementation of a statistical model, which can be
successfully applied for compressing of a sequence of binary digits with
behavior close to random.
\end{abstract}

\section{Introduction}
  It's well known that in most cases Huffman encoding is not the optimal
encoding of a stream of characters of an alphabet associatd with a
probabilistic model \cite{nelson2}. Although a Huffman encoded sequence
of data can look like ``random'', most likely it is not a real noise- like
the one can get from \verb|/dev/random| in some UNIX(-like) systems. In
the present article we give a ``proof of concept'' implementation of
a statistical model capable of detecting the deviation of the data
being encoded from the notion of ``real noise''.
\section{The Model}
  We break the input data into consequtive blocks $B_1,B_2,B_3,\ldots$
of size $N$ bits each. Each block $B_i$ is being encoded in the following way:
\begin{enumerate}
\item{}Count the number of 1's $k_i$ in $B_i$.
\item{}Let $c_1=k_i$, $c_0=N-k_i$. The probability to encounter a 0 in
the block is $c_0/N$, and for 1 is $c_1/N$
\item{}Encode a bit of the block with the calculated probabilities.
Depending on its value we decrease
either $c_0$ or $c_1$.
\item{}Proceed with all the bits of the block. Note: obviously the last
bit won't need to be encoded, because we know its value with probability
1. Anyway, a decently implemented arithmetic encoder should not encode
anything when given probability 1.
\end{enumerate}
The sequence $k_i$ is also a data stream, which needs to be encoded
for obvious reasons. The most reasonable scheme for its encoding is
0-order adaptive statistical model + arithmetic encoding. For ``real noise''
$k_i$ is supposed to have a binomial distribution:
\begin{equation}\label{bin1}
P(k)=\frac{N!}{2^Nk!(N-k)!},
\end{equation}
and ``slightly compressible data'' (``not-so-real noise'') could be
expected to have some deviation from this distribution. We make a
plausible conjecture that data whose $k_i$ is deviating from
\eqref{bin1}, is compressible. An example of such deviation is shown in
the figure.

\begin{figure}[h]
\begin{center}
\epsfxsize=0.5\textwidth
\epsfbox{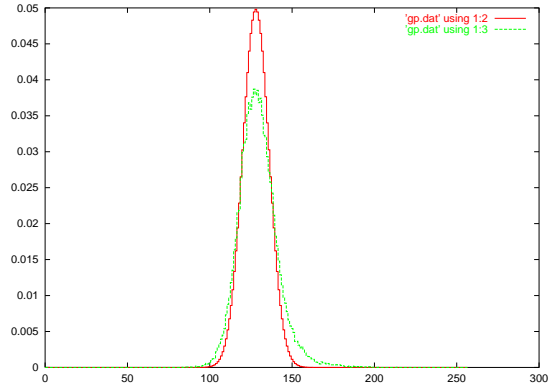}
\end{center}
\caption{The distribution of $k_i$ for the file
xscreensaver-4.00.tar.gz compared to binomial distribution}
\end{figure}

\section{Source Code}
We provide the C++ source code for the implementation of the algorithm
described above. The remaining source modules- implementation of arithmetic
encoding, front end, error handling etc., necessary to compile a working
program, are not included, but easy to implement \cite{nelson2}.
\begin{verbatim}
#include "bitclasses.h"
/*#undef TRACE_*/




long filelen(FILE*f)
{
  long len,pos;
  pos=ftell(f);
  fseek(f,0L,SEEK_END);
  len=ftell(f);
  fseek(f,pos,SEEK_SET);
  return len;
}

static long double facto(int k){
  long double res;
  res=1;
  while(k>1)res*=k--;
  return res;
}
class chistomodel{
public:
  int n;
  int *tab;
  float pro(int k)
  {
  return facto(n)/(facto(k)*facto(n-k)*2.0);
  }
  ~chistomodel(){
  if(tab)
    delete tab;
  }
  chistomodel(){tab=NULL;}
  chistomodel(int n)
  {tab=NULL;init(n);}
  void init(int n){
  }
};

#if 1
#define B_SIZE 32
#else
#define B_SIZE 32
#endif
#define BITSINBLOCK (B_SIZE*8)
class ccmodel{
public:
  int tab[BITSINBLOCK+2];
  unsigned short scale(){return tab[BITSINBLOCK+1];}
  ccmodel(){init();}
  void init()
  {
  int i;
  for(i=0;i<=BITSINBLOCK+1;i++)
    tab[i]=i;
  }
  void update(int c)
  {
  int i;
  if(tab[BITSINBLOCK+1]>=MAXSCALE)
    rescale();
  for(i=c+1;i<=BITSINBLOCK+1;i++)
    tab[i]++;
  }
  void rescale()
  {
  unsigned weights[BITSINBLOCK+1];
  int i;
  for(i=0;i<BITSINBLOCK+1;i++)
    {
    weights[i]=(tab[i+1]-tab[i])>>1;
    if(!weights[i])
      weights[i]=1;
    }
  for(i=0;i<BITSINBLOCK+1;i++)
    tab[i+1]=tab[i]+weights[i];
  }
  void chartosymb(int c,csymbol*s)
  {
  s->low_count=tab[c];
  s->high_count=tab[c+1];
  s->scale=tab[BITSINBLOCK+1];
  }
  int counttochar(short int count,csymbol*s)
  {
  int l,h,m;
  l=0;
  h=BITSINBLOCK+1;
  while(h-l>1)
    {
    m=(h+l)>>1;
    if(tab[m]<=count)
      l=m;
    else
      h=m;
    }
  s->low_count=tab[l];
  s->high_count=tab[l+1];
  s->scale=tab[BITSINBLOCK+1];
  return l;
  }
};
extern int bittable[];
unsigned char block[B_SIZE];
int br;
ccmodel countmodel;
void decode_block(carithmeticdecoder&ari)
{
  int s,count;
  int bit1cnt,bit0cnt;
  unsigned mask;
  csymbol symb;
  count=ari.get_current_count(countmodel.scale());
  bit1cnt=countmodel.counttochar(count,&symb);
  ari.remove_symbol_from_stream(symb.low_count,
                symb.high_count,
                symb.scale);
  bit0cnt=br*8-bit1cnt;
  countmodel.update(bit1cnt);
  for(s=0;s<br;s++)
  {
    block[s]=0;
    for(mask=0x80;mask;mask>>=1)
    {
      count=ari.get_current_count(bit0cnt+bit1cnt);
      if(count<bit0cnt)
      {
        ari.remove_symbol_from_stream(0,
                      bit0cnt,
                      bit0cnt+bit1cnt);
        bit0cnt--;
      }
      else
      {
        block[s]|=mask;
        ari.remove_symbol_from_stream(bit0cnt,
                      bit0cnt+bit1cnt,
                      bit0cnt+bit1cnt);
        bit1cnt--;
      }
    }
  }
}
void encode_block(carithmeticencoder&ari)
{
  int i;
  int bit1cnt=0,bit0cnt;
  int mask;
  csymbol symb;
  for(i=0;i<br;i++)
  bit1cnt+=bittable[block[i]];
  bit0cnt=br*8-bit1cnt;
#if 1
#if 1
  fprintf(stdout,"%d\n",bit1cnt);
#else
  fputc(bit1cnt,stdout);
#endif
#endif
#if 0
  if(bit1cnt>255)
  fprintf(stderr,"$%d\n",bit1cnt);
#endif
  countmodel.chartosymb(bit1cnt,&symb);
  ari.encode_symbol(symb.low_count,
          symb.high_count,
          symb.scale);
  countmodel.update(bit1cnt);
  for(i=0;i<br;i++)
  {
    for(mask=0x80;mask;mask>>=1)
    if(block[i]&mask)
      {
      symb.low_count=bit0cnt;
      symb.high_count=bit0cnt+bit1cnt;
      symb.scale=bit0cnt+bit1cnt;
      ari.encode_symbol(symb.low_count,
                symb.high_count,
                symb.scale);
      bit1cnt--;
      }
    else
      {
      symb.low_count=0;
      symb.high_count=bit0cnt;
      symb.scale=bit0cnt+bit1cnt;
      ari.encode_symbol(symb.low_count,
                symb.high_count,
                symb.scale);
      bit0cnt--;
      }
  }
}
void do_compress()
{
  long length,l;
  coutbitstream outstr;
  carithmeticencoder ari;
  l=length=filelen(infile);
  fwrite(&length,sizeof(length),1,outfile);
  outstr.init(outfile);
  ari.init(&outstr);
  while(1)
  {
    br=fread(block,1,B_SIZE,infile);
    l-=br;
    encode_block(ari);
    if((br<B_SIZE)||(l<1))
    break;
  }
  ari.flush();
  outstr.flush();
}
void do_decompress()
{
  long length,l;
  cinbitstream instr;
  carithmeticdecoder ari;
  csymbol symb;
  fread(&length,sizeof(length),1,infile);
  l=length;
  instr.init(infile);
  ari.init(&instr);
  while(l>0)
  {
    br=l<B_SIZE?l:B_SIZE;
    l-=br;
    decode_block(ari);
    fwrite(block,1,br,outfile);
  }
}

int bittable[]={
0,1,1,2,1,2,2,3,1,2,2,3,2,3,3,4,1,2,2,3,2,3,3,4,2,3,3,4,3,4,4,5,1,2,2,3,2,3,3,
4,2,3,3,4,3,4,4,5,2,3,3,4,3,4,4,5,3,4,4,5,4,5,5,6,1,2,2,3,2,3,3,4,2,3,3,4,3,4,
4,5,2,3,3,4,3,4,4,5,3,4,4,5,4,5,5,6,2,3,3,4,3,4,4,5,3,4,4,5,4,5,5,6,3,4,4,5,4,
5,5,6,4,5,5,6,5,6,6,7,1,2,2,3,2,3,3,4,2,3,3,4,3,4,4,5,2,3,3,4,3,4,4,5,3,4,4,5,
4,5,5,6,2,3,3,4,3,4,4,5,3,4,4,5,4,5,5,6,3,4,4,5,4,5,5,6,4,5,5,6,5,6,6,7,2,3,3,
4,3,4,4,5,3,4,4,5,4,5,5,6,3,4,4,5,4,5,5,6,4,5,5,6,5,6,6,7,3,4,4,5,4,5,5,6,4,5,
5,6,5,6,6,7,4,5,5,6,5,6,6,7,5,6,6,7,6,7,7,8
};
\end{verbatim}

\end{document}